\documentclass{article}
\usepackage{latexsym}
\usepackage{graphics}
\usepackage{color}
\usepackage{amsmath, amsthm, amssymb}
\newtheorem{proposition}{Proposition}[section]
\newtheorem{theorem}{Theorem}[section]

\newtheorem{corollary}{Corollary}[section]

\newtheorem{definition}{Definition}[section]
\newtheorem{assumption}{Assumption}[section]
\newtheorem{property}{Property}[section]

\begin{document}
\title{On naked singularities 
and the collapse of self-gravitating Higgs
fields}
\author{Mihalis Dafermos}
\maketitle
\begin{abstract}
We consider the problem of collapse of a self-gravitating
Higgs field, with potential bounded below by a (possibly negative)
constant. The behaviour at infinity may be either asymptotically
flat or asymptotically AdS. 
This problem has recently received much attention
as a source for possible violations of weak cosmic censorship
in string theory. In this paper, we prove under spherical symmetry
that ``first singularities'' arising in the
non-trapped region must necessarily emanate from the centre. In particular, this 
excludes the formation of a certain type of naked singularity which was
recently conjectured to occur.
\end{abstract}
A fundamental open problem in classical general relativity is
the problem of weak cosmic censorship, the conjecture
that generic asymptotically flat initial data lead to a Cauchy development with
a complete null infinity. 

In the case of a spherically symmetric scalar field with vanishing potential,
the conjecture was resolved in the affirmative by Christodoulou~\cite{chr:ins}.
Previously, Christodoulou had
proven~\cite{chr:ns} that the caveat ``generic'' was necessary, by constructing
explicit examples of naked singularities forming from the collapse of regular data.
These naked singularities emanate from the centre.

It has been suggested recently~\cite{hhm:gccv, gh:cns}
that the situation changes drastically for self-gravitating Higgs
fields with potentials which can take negative values.
Such ``non-classical''\footnote{Higgs fields with negative
potential violate the dominant energy condition.} matter 
can be motivated by considerations arising from string theory.
In~\cite{hhm:gccv, gh:cns}, the authors
advance certain heuristic arguments to show that 
naked singularities not emanating from the centre can arise
and in fact \emph{will} arise for an \emph{open} set of initial data.

In this paper, we will prove a simple estimate for self-gravitating
Higgs fields with potential bounded below by a (possibly negative) constant.
In the evolutionary context, this estimate shows 
that ``first singularities'' arising from the \emph{nontrapped
region}\footnote{This is the region of spacetime such that the outgoing expansion
of the group orbit spheres is nonnegative, whereas the ingoing expansion is
negative.}
must necessarily emanate from the center. Since 
the past of infinity (in either the asymptotically
flat or asymptotically AdS setting) must lie in the nontrapped
region\footnote{This fact only depends on the null convergence
condition, which is still satisfied here.}, 
this implies, in particular, that the naked singularities
of~\cite{gh:cns}
can in fact never arise.

Finally, in the case where the potential is 
in fact \emph{non-negative}, then the results
of of this paper show that Higgs fields satisfy the assumptions
of~\cite{md:sssts}. In particular, the existence of a single trapped
surface is sufficient to deduce the completeness of null infinity.
The reader is referred to~\cite{md:sssts}.

\section{Basic assumptions}
A self-gravitating scalar
field with potential is described by
a four-dimensional spacetime $(\mathcal{M},g)$
and a function $\phi$ on $\mathcal{M}$ satisfying:
\begin{equation}
\label{Einstein-xs}
R_{\mu\nu}-\frac12g_{\mu\nu}R=2T_{\mu\nu}
\end{equation}
\begin{equation}
\label{wave-xs}
\phi_{;\mu}^{;\mu}=V'(\phi)
\end{equation}
\begin{equation}
\label{em-xs}
T_{\mu\nu}=\phi_{;\mu}\phi_{;\nu}-\frac12g_{\mu\nu}\phi_{;\alpha}\phi^{;\alpha}
-g_{\mu\nu}V(\phi),
\end{equation}
where $V$ is a continuously differentiable function of its argument:
\begin{equation}
\label{duvamiko2}
V\in C^1({\bf R}).
\end{equation}

This paper will concern such spacetimes which are \emph{spherically symmetric},
and \emph{evolutionary}. 

By \emph{spherically symmetric}, we here mean
that the group
$SO(3)$ acts by isometry on $\mathcal{M}$, and preserves
$\phi$. 
Actually, we will require something a  little stronger,
namely that
\[
\mathcal{Q}^+=
\mathcal{M}/SO(3)
\] 
inherit from 
$g$ the structure of a $1+1$-dimensional 
time-oriented Lorentzian manifold with boundary,
with metric $\bar{g}$,
such that 
\[
g=\bar{g}+r^2\gamma,
\]
where $\gamma$ here denotes the standard metric on $S^2$,
and $r$ is a nonnegative function on $\mathcal{Q}^+$,
the so-called \emph{area-radius} function.
Since $\phi$ is constant on group orbits, it
descends to a function defined on $\mathcal{Q}^+$.

By \emph{evolutionary}, we mean that the spacetime
is to have those properties one would require from the
``unique maximal development'' of an appropriate initial or initial-boundary
value problem for $(\ref{Einstein-xs})$--$(\ref{em-xs})$.
We have two such problems in mind:
\begin{enumerate}
\item
Initial data are defined on a complete asymptotically flat
hypersurface, leading to an asymptotically flat maximal Cauchy development,
or alternatively 
\item
Initial data are prescribed on an asymptotically AdS hypersurface,
supplemented with boundary conditions at infinity, leading to an
asymptotically AdS spacetime.
\end{enumerate}
Only in the former case, has the construction
of such a maximal development been rigorously carried out.\footnote{In particular,
in this case, the assumption of spherical symmetry described above
can be retrieved from an analogous assumption on initial data.}
Nevertheless, the properties to be listed here must be satisfied for \emph{any}
reasonable notion of such an object. 

We proceed here to list these properties. For
the purposes of mathematical clarity, the reader can chose to consider
the ``properties'' listed in this section as a priori \emph{assumptions}.

\subsection{Geometry of $\mathcal{Q}^+$}
\label{gewmetria}
The boundary\footnote{Recall that $\mathcal{Q}^+$ is a manifold with boundary.
The boundary is not quite a $1$-dimensional manifold, but a piecewise
regular curve.} 
of $\mathcal{Q}^+$ will consist of 
$\Gamma\cup S$, where $\Gamma$ is a connected timelike curve, 
and $S$ is a connected spacelike curve, and $\Gamma \cap S$ is a 
single point,
and such that $r(q)=0$ iff $q\in \Gamma$.
We will call $\Gamma$ the \emph{centre}.
Let $L$ denote a null parallel vector field along
$\Gamma\cup S$ pointing towards $\mathcal{Q}^+$.  
$\mathcal{Q}^+$ will be foliated by the family
of null geodesics defined by $L$. We will call these geodesics ``outgoing''.

To discuss the behaviour of the boundary ``at infinity'', it is convenient
to introduce what are essentially Penrose diagrams.
Our quotient spacetime $\mathcal{Q}^+$
will admit a $C^3$ map into a bounded subset of 
$2$-dimensional Minkowski space which preserves the causal structure.
We henceforth identify $\mathcal{Q}^+$ with its image under such a map.
Standard null coordinates $u$ and $v$ provide a global coordinate chart
on $\mathcal{Q}^+$, and the metric $\bar{g}$
can be written $-\Omega^2dudv$. Let
$u$ be such that constant-$u$ curves are ``outgoing'', as defined above.

We assume that there exists a non-empty causal curve 
$\mathcal{I}\subset\overline{\mathcal{Q}^+}\setminus
\mathcal{Q}^+$,\footnote{i.e., the 
set $\overline{\mathcal{Q}^+}\setminus\mathcal{Q}$
denotes the portion of the boundary of $\overline{\mathcal{Q}^+}$ as as subset of
${\bf R}^2$, which is not part of its boundary, in the sense of manifolds-with-boundary.}
and that
$\mathcal{Q}^+$ is foliated by connected constant
$v$-curves with past endpoint on $S$ or past end-limit point
on $\mathcal{I}$. We call such null geodesics ``ingoing''. 
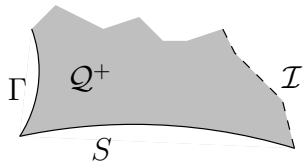
\[
\begin{picture}(0,0)%
\includegraphics{A.pstex}%
\end{picture}%
\setlength{\unitlength}{3947sp}%
\begingroup\makeatletter\ifx\SetFigFont\undefined%
\gdef\SetFigFont#1#2#3#4#5{%
  \reset@font\fontsize{#1}{#2pt}%
  \fontfamily{#3}\fontseries{#4}\fontshape{#5}%
  \selectfont}%
\fi\endgroup%
\begin{picture}(1812,1025)(3001,-4010)
\put(3376,-3511){\makebox(0,0)[lb]{\smash{\SetFigFont{12}{14.4}{\rmdefault}{\mddefault}{\updefault}{\color[rgb]{0,0,0}$\mathcal{Q}^+$}%
}}}
\put(3001,-3586){\makebox(0,0)[lb]{\smash{\SetFigFont{12}{14.4}{\rmdefault}{\mddefault}{\updefault}{\color[rgb]{0,0,0}$\Gamma$}%
}}}
\put(3526,-3961){\makebox(0,0)[lb]{\smash{\SetFigFont{12}{14.4}{\rmdefault}{\mddefault}{\updefault}{\color[rgb]{0,0,0}$S$}%
}}}
\put(4726,-3511){\makebox(0,0)[lb]{\smash{\SetFigFont{12}{14.4}{\rmdefault}{\mddefault}{\updefault}{\color[rgb]{0,0,0}$\mathcal{I}$}%
}}}
\end{picture}

\]

\subsection{Reduced equations}
\label{regsec}
From the equations $(\ref{Einstein-xs})$--$(\ref{em-xs})$
we derive
\begin{equation}
\label{evolution1}
\partial_u\partial_v r
=-\frac1r\partial_ur\partial_vr-\frac1{4r}\Omega^2+\frac12r\Omega^2V(\phi),
\end{equation}
\begin{equation}
\label{evolution2}
\partial_u\partial_v \log \Omega^2=
\frac14\Omega^2 r^{-2}+r^{-2}\partial_ur\partial_vr
-\partial_u\phi\partial_v\phi,
\end{equation}
\begin{equation}
\label{evolution3}
\partial_u\partial_v \phi=-r^{-1}\partial_u\phi\partial_vr-
r^{-1}\partial_v\phi\partial_ur
-\frac14\Omega^2V'(\phi),
\end{equation}
\begin{equation}
\label{constraint1}
\partial_u(\Omega^{-2}\partial_ur)=-r\Omega^{-2}(\partial_u\phi)^2,
\end{equation}
\begin{equation}
\label{constraint2}
\partial_v(\Omega^{-2}\partial_vr)=-r\Omega^{-2}(\partial_v\phi)^2.
\end{equation}
We assume here that $r$, $\Omega$, and $\phi$ are $C^2$, and that
these equations hold pointwise.

\subsection{Local existence and extendibility criterion}
Evolutionary spacetimes are constructed by a local existence and uniqueness
theorem in an appropriately defined function space. One first shows
that there exists a non-empty ``development'' of initial data. 
Then, using the domain of dependence property,
it is straightforward to show that there exists a 
unique \emph{maximal} development.

Our embedding of $\mathcal{Q}^+$ into ${\bf R}^{1+1}$
allows us to talk about its future boundary as a subset of ${\bf R}^{1+1}$.
In what follows
let $\overline{\mathcal{Q}^+}$ denote the closure of $\mathcal{Q}^+$ 
\emph{in the topology of ${\bf R}^{1+1}$}, and similarly
$\overline{\mathcal{R}}$, etc., and let $J^-$, $J^+$, $D^+$, etc., refer to the
causal structure of the ${\bf R}^{1+1}$.
In this section, we shall formulate a criterion for
a point $p\in \overline{\mathcal{Q}^+}\setminus(\overline{\mathcal{I}}
\cup \overline{\Gamma})$
to be, in a suitable
sense, a ``first singularity''.
Since we are interested in considerations
away from the centre, the following double null local existence
result will be sufficient for our purposes:
\begin{proposition}
\label{local}
Let $\Omega$, $r$, and $\phi$ be functions defined
on $X=[0,d]\times \{0\}\cup \{0\}\times[0,d]$. Let $k\ge0$, and assume 
$r>0$ is $C^{k+2}(u)$
on $[0,d]\times \{0\}$
and $C^{k+2}(v)$ on $\{0\}\times[0,d]$,
assume that $\Omega$ and $\phi$ are $C^{k+1}(u)$ on
$[0,d]\times\{0\}$ and $C^{k+1}(v)$
on $\{0\}\times[0,d]$, and assume that $V$ is a $C^{k+1}$ function
of its argument.
Suppose that equations $(\ref{constraint1})$, $(\ref{constraint2})$
hold initially on $[0,d]\times\{0\}$ and $\{0\}\times[0,d]$, respectively.
Let $|\Omega|_{n,u}$ denote the $C^{n}(u)$ norm of $\Omega$ on
$[0,d]\times \{0\}$, $|\Omega|_{n,v}$ the $C^{n}(v)$ norm of $\Omega$
on $\{0\}\times[0,d]$, etc.
Define
\[
N=\sup\{|\Omega|_{1,u}, |\Omega|_{1,v},
|\Omega^{-1}|_0, |r|_{2,u}, |r|_{2,v}, |r|^{-1}_{0},
|\phi|_{1,u}, |\phi|_{1,v}\}.
\]
Then there exists a $\delta$,
depending only on $N$,
and a $C^{k+2}$ function (unique among $C^2$ functions)
$r$ and $C^{k+1}$ functions (unique among $C^1$ functions)
$\Omega$ and $\phi$,
satisfying equations $(\ref{evolution1})$--$(\ref{constraint2})$ 
in $[0,\delta^*]\times[0,\delta^*]$, where
$\delta^*=\min\{d,\delta\}$, such that
the restriction of these functions
to $[0,d]\times\{0\}\cup\{0\}\times[0,d]$
is as prescribed.
\end{proposition}

The proof can be obtained by standard methods and is omitted.
To describe the characterization of (part of) the boundary of $\mathcal{Q}^+$
that this leads to, first let us introduce some terminology.

\begin{definition}
\label{orismos}
Let $p\in\overline{\mathcal{Q}^+}$. The indecomposable past subset
$J^-(p)\cap\mathcal{Q}^+\subset\mathcal{Q}^+$ 
is said to be \emph{eventually compactly generated}
if there exists a compact subset
$X\subset \mathcal{Q}^+$ such that
\begin{equation}
\label{astnr}
J^{-}(p)\subset D^+(X)\cup J^-(X).
\end{equation}
\end{definition}
\begin{definition}
A point $p\in\overline{\mathcal{Q}^+}\setminus\mathcal{Q}^+$ is said
to be a \emph{first singularity} if $J^-(p)\cap\mathcal{Q}^+$  is
eventually
compactly generated and 
if any eventually compactly generated indecomposable proper subset
of $J^-(p)\cap\mathcal{Q}^+$ is of the form $J^-(q)$ for a
$q\in\mathcal{Q}^+$.
\end{definition}

In particular, in view of the geometry of $\mathcal{Q}^{+}$,
as described in Section~\ref{gewmetria},
setting
$p=(u_s,v_s)$, then if $p$ is a first
singularity and if $p\not\in \overline\Gamma$, then this
implies that there exists an $\epsilon>0$
such that for any $u_s>u_\epsilon>u_s-\epsilon$,
$v_s>v_\epsilon>v_s-\epsilon$, the compact set
\[
X=\{u_\epsilon\}\times[v_\epsilon,v_s]\cup [u_\epsilon,u_s]\times\{v^*\}
\]
satisfies $X\subset \mathcal{Q}^+\setminus\Gamma$, 
and we have
\[
[u_\epsilon,v_\epsilon]\times[u_s,v_s]=D^+(X)=J^{-}(p)\cap D^+(X),
\]
and 
\[
D^+(X)\cap\mathcal{Q}^+=D^+(X)\setminus\{p\}.
\]

Given a subset $Y\subset\mathcal{Q}^+\setminus\Gamma$,
define
\[
N(Y)=\sup\{|\Omega|_{1}, 
|\Omega^{-1}|_0, |r|_{2}, |r|^{-1}_{0},
|\phi|_{1}\},
\]
where, for $f$ defined on $\mathcal{Q}^+$, $|f|_k$ denotes the
restriction of the $C^k$ norm to $Y$.

In the evolutionary context, Proposition~\ref{local} easily gives
the following \emph{extension criterion}
\begin{property}
\label{epekt}
Let $p\in\overline{\mathcal{Q}^+}\setminus\overline\Gamma$ be a first singularity.
Then for any compact
$X\subset\mathcal{Q}^+\setminus\Gamma$ satisfying $(\ref{astnr})$, we have
\[
N(D^+(X)\setminus\{p\})=\infty.
\]
\end{property}

\section{$\mathcal{R}$, $\mathcal{T}$, and $\mathcal{A}$}
Let us introduce the notation:
\begin{equation}
\label{ruqu}
\nu=\partial_u{r},
\end{equation}
\begin{equation}
\label{rvqu}
\lambda=\partial_v{r}.
\end{equation}
We define the \emph{regular region} 
\[
\mathcal{R}=\{q\in\mathcal{Q}^+:\lambda(q)>0, \nu(q)<0\},
\] 
the \emph{trapped region} 
\[
\mathcal{T}=\{q\in\mathcal{Q}^+:\lambda(q)<0, \nu(q)<0\}, 
\]
and the \emph{marginally trapped region}
\[
\mathcal{A}=\{q\in\mathcal{Q}^+:\lambda(q)=0, \nu(q)<0\}.
\]
This notation derives from~\cite{chr:sgrf}.
We call $\mathcal{R}\cup\mathcal{A}$ the  \emph{non-trapped} region.

In what follows we will now always assume
that the following hold:
\begin{assumption}
\label{no-trapped}
We have 
\[
\partial_ur < 0
\]
along $S$.
\end{assumption}
\begin{property} 
\label{inftystr}
For all $q\in\mathcal{I}$,  for all
$p\in J^+(q)\cap\mathcal{Q}^+\setminus I^+(q)$, and for all $R>0$,
there exists a $p^*\in J^+(q)\cap J^-(p)\cap\mathcal{Q}^+\cap \setminus I^+(q)$
such that $r(p^*)\ge R$. 
Similarly, for all $q\in\mathcal{I}$, all
$p\in J^-(q)\cap\mathcal{Q}^+\setminus I^-(q)$, and all $R>0$,
there exists a $p^*\in J^-(q)\cap J^+(p)\cap\mathcal{Q}^+\cap \setminus I^-(q)$
such that $r(p^*)\ge R$.
\end{property}

Assumption~\ref{no-trapped} is motivated in~\cite{chr:sgrf}.  
Property~\ref{inftystr}
is perhaps the weakest possible notion that $\mathcal{I}$ is actually
``at'' infinity. 
Note that the set defined by $\mathcal{I}^+$
in~\cite{md:sssts}  for the asymptotically flat case clearly  satisfies 
Property~\ref{inftystr}.

We have
\begin{proposition}
\label{prosnmo-}
$\nu<0$ everywhere,
i.e.~$\mathcal{Q}^{+}=\mathcal{A}\cup\mathcal{R}\cup\mathcal{T}$.
\end{proposition}

\noindent\emph{Proof.} 
The previous assumption, together with the global structure
of $\mathcal{Q}^+$, as described in Section~\ref{gewmetria},
implies that if $(u,v)\in\mathcal{Q}^+$, then there
exists a $u^*<u$ such that $\nu(u^*,v)<0$. The proposition
now follows immediately after integration of $(\ref{constraint1})$ along $[u^*,u]\times\{v\}$.
$\Box$.

Similarly from $(\ref{constraint2})$, we immediately obtain,
\begin{proposition}
\label{inview}
We have
\[
J^-(\mathcal{I})\cap\mathcal{Q}^+\subset \mathcal{R}
\]
and
\[
\overline{J^-(\mathcal{I})}\cap\mathcal{Q}^+\subset 
\mathcal{R}\cup\mathcal{A}.
\]
\end{proposition} 
In particular,
\begin{corollary}
\label{allo}
If $\mathcal{T}\ne\emptyset$, then 
$\mathcal{Q}^+\setminus J^-(\mathcal{I})\ne\emptyset$.
\end{corollary}

\section{The main theorem}
Let us introduce one final assumption:
\begin{assumption}
\label{bndbelow}
There exists a $C$ such that:
\begin{equation}
\label{duvamiko1}
V(x)\ge -C.
\end{equation}
\end{assumption}

The main theorem of this paper is
\begin{theorem}
\label{kurio}
Let $p\in\overline{\mathcal{Q}^+}\setminus\mathcal{Q}^+$ 
be a first singularity. Then either
\begin{equation}
\label{suv9nkn1}
p\in\overline{\Gamma}\setminus\Gamma
\end{equation}
or
\begin{equation}
\label{suv9nkn2}
J^-(p)\cap\mathcal{Q}^+\cap D^+(X)\cap\mathcal{T}\neq\emptyset,
\end{equation}
for all compact $X$ satisfying $(\ref{astnr})$.
\end{theorem}

\section{Proof of the main theorem}
It is equivalent to prove the following:
Let $p\in\overline{\mathcal{Q}^+}\setminus\overline{\Gamma}$ be
such that $J^-(p)\cap\mathcal{Q}^+$ is eventually compactly generated,
and such that any compactly generated indecomposable subset
$J^-(q)\cap\mathcal{Q}^+\subset J^-(p)\cap\mathcal{Q}^+$ satisfies
$q\in\mathcal{Q}^+$. Then $p\in \mathcal{R}\cup\mathcal{A}$.

Choose $\epsilon$, $X$, $u_\epsilon$, $v_\epsilon$, as in the statement
following the definition of a first singularity, so that in addition
$D^+(X)\setminus\{p\}\subset\mathcal{R}\cup\mathcal{A}$.
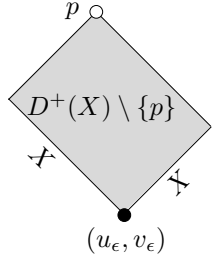
\[
\begin{picture}(0,0)%
\includegraphics{e3artnsnN.pstex}%
\end{picture}%
\setlength{\unitlength}{3158sp}%
\begingroup\makeatletter\ifx\SetFigFont\undefined%
\gdef\SetFigFont#1#2#3#4#5{%
  \reset@font\fontsize{#1}{#2pt}%
  \fontfamily{#3}\fontseries{#4}\fontshape{#5}%
  \selectfont}%
\fi\endgroup%
\begin{picture}(1617,2047)(5524,-5872)
\put(5985,-3981){\makebox(0,0)[lb]{\smash{\SetFigFont{10}{12.0}{\familydefault}{\mddefault}{\updefault}{\color[rgb]{0,0,0}$p$}%
}}}
\put(6853,-5420){\rotatebox{45.0}{\makebox(0,0)[lb]{\smash{\SetFigFont{10}{12.0}{\familydefault}{\mddefault}{\updefault}{\color[rgb]{0,0,0}$X$}%
}}}}
\put(5683,-5091){\rotatebox{315.0}{\makebox(0,0)[lb]{\smash{\SetFigFont{10}{12.0}{\familydefault}{\mddefault}{\updefault}{\color[rgb]{0,0,0}$X$}%
}}}}
\put(6144,-5814){\makebox(0,0)[lb]{\smash{\SetFigFont{10}{12.0}{\familydefault}{\mddefault}{\updefault}{\color[rgb]{0,0,0}$(u_\epsilon,v_\epsilon)$}%
}}}
\put(5680,-4771){\makebox(0,0)[lb]{\smash{\SetFigFont{10}{12.0}{\familydefault}{\mddefault}{\updefault}{\color[rgb]{0,0,0}$D^+(X)\setminus\{p\}$}%
}}}
\end{picture}

\]
Let us introduce the notation:
\begin{equation}
\label{uphi}
\zeta=r\partial_u\phi,
\end{equation}
\begin{equation}
\label{vphi}
\theta=r\partial_v\phi,
\end{equation}
and the quantity
(in view of Proposition~\ref{prosnmo-}) 
\[
\kappa=-\frac14\Omega^2\nu^{-1}.
\]
We clearly have
\[
\kappa>0.
\] 

Key to our proof is the \emph{Hawking mass} function, defined by
\[
m=\frac r2\left(1+4\Omega^{-2}\partial_ur\partial_vr\right).
\] 
It is convenient to define also
the so-called
\emph{mass aspect function} 
\[
\mu=\frac{2m}r.
\]

By compactness of $X$, and the regularity assumptions
of Section~\ref{regsec}, it follows
that $r$, $\kappa$,
$\theta$, $\zeta$, $\phi$, $\lambda$, $\nu$, $m$, 
$\partial_u\Omega$, $\partial_v\Omega$, $\partial_v\lambda$, $\partial_u\nu$
are uniformly bounded
above and below on $X$:
\begin{equation}
\label{yeniyildiz2}
0<r_0\le r\le R,
\end{equation}
\[
0\le \lambda\le \Lambda,
\]
\[
0>\nu_0\ge\nu\ge -N,
\]
\[
|\phi|\le P,
\]
\[
|\theta|\le\Theta,
\]
\[
|\zeta|\le Z,
\]
\begin{equation}
\label{maza-fragma}
|m|\le M,
\end{equation}
\begin{equation}
\label{kai-auto}
0<\kappa\le K,
\end{equation}
\begin{equation}
\label{para1}
|\partial_u\Omega|\le H,
\end{equation}
\begin{equation}
\label{para2}
|\partial_v\Omega|\le H,
\end{equation}
\begin{equation}
\label{para3}
|\partial_u\nu|\le H,
\end{equation}
\begin{equation}
\label{para4}
|\partial_v\lambda|\le H.
\end{equation}
In view of Property~\ref{epekt}, it will be enough to show that uniform bounds similar
to the above hold throughout $D^+(X)\setminus\{p\}$.

The importance of the Hawking mass derives from the following
identities (see~\cite{chr:sgrf}):
\begin{equation}
\label{puqu}
\partial_u m=\frac{1}{2}(1-\mu)\left(\frac\zeta\nu\right)^2\nu+
r^2V(\phi)\nu,
\end{equation}
\begin{equation}
\label{pvqu}
\partial_v m=\frac{1}{2}\kappa^{-1}\theta^2
+r^2V(\phi)\lambda.
\end{equation}
To understand the nature of the terms in $(\ref{puqu})$ and $(\ref{pvqu})$,
 note first that 
\[
(1-\mu)\kappa=\lambda,
\]
and thus
\begin{equation}
\label{yeniyildiz}
1-\mu\ge0
\end{equation}
on $\mathcal{R}\cup\mathcal{A}$.
The equation $(\ref{constraint1})$ yields
\begin{equation}
\label{fdb1}
\partial_u\kappa=
\frac1{r}\left(\frac\zeta\nu\right)^2\nu\kappa.
\end{equation}
From $(\ref{fdb1})$, and Proposition~\ref{prosnmo-}, it follows
that the bound $(\ref{kai-auto})$ holds throught
$D^+(X)\setminus\{p\}$.

The idea now of the proof is as follows:
As we shall see momentarily, the fact that $D^+(X)\setminus\{p\}\subset
\mathcal{A}\cup\mathcal{R}$ immediately yields 
that the bounds $(\ref{yeniyildiz2})$ are preserved.
If $V$ were nonnegative, then the signs in equations $(\ref{puqu})$ and
$(\ref{pvqu})$ would immediately yield that $(\ref{maza-fragma})$ is preserved.
In view of the bounds $(\ref{yeniyildiz2})$ on $r$ and 
$(\ref{kai-auto})$ on $\kappa$, integration of $(\ref{pvqu})$ in $v$
would yield an $L^2(v)$ bound for $\partial_v\phi$. From this, 
bounds on all other quantities
would follow in a straightforward manner.

In our case, $V$ is of course not nonnegative. It turns out, however,
that in view of Assumption~\ref{bndbelow}, we can 
still derive estimates on $m$, since we can control the integral
of the term with the ``wrong'' sign from $(\ref{duvamiko1})$ and the bounds 
$(\ref{yeniyildiz2})$ on $r$.\footnote{It is clear then that $(\ref{duvamiko1})$ is
essential in this argument; indeed, if $(\ref{duvamiko1})$ is violated,
there is no reason to believe that this theorem holds.}
In view of the triangle inequality, we can still obtain
an $L^2$ bound on $\partial_v\phi$ from our bounds $(\ref{yeniyildiz})$ on $r$ 
and the bounds just obtained for $m$,
upon integration of $(\ref{pvqu})$. We then continue as before.


Now for the details:
Given any point $(u^*,v^*)\in D^+(X)\setminus\{p\}$, 
the null curves
$u=u^*$ and $v=v^*$ both intersect $X$: 
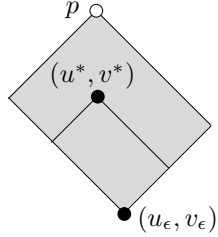
\[
\begin{picture}(0,0)%
\includegraphics{e3artnsnN2.pstex}%
\end{picture}%
\setlength{\unitlength}{3158sp}%
\begingroup\makeatletter\ifx\SetFigFont\undefined%
\gdef\SetFigFont#1#2#3#4#5{%
  \reset@font\fontsize{#1}{#2pt}%
  \fontfamily{#3}\fontseries{#4}\fontshape{#5}%
  \selectfont}%
\fi\endgroup%
\begin{picture}(1617,1882)(5524,-5707)
\put(5985,-3981){\makebox(0,0)[lb]{\smash{\SetFigFont{10}{12.0}{\familydefault}{\mddefault}{\updefault}{\color[rgb]{0,0,0}$p$}%
}}}
\put(6549,-5649){\makebox(0,0)[lb]{\smash{\SetFigFont{10}{12.0}{\familydefault}{\mddefault}{\updefault}{\color[rgb]{0,0,0}$(u_\epsilon,v_\epsilon)$}%
}}}
\put(5859,-4524){\makebox(0,0)[lb]{\smash{\SetFigFont{10}{12.0}{\familydefault}{\mddefault}{\updefault}{\color[rgb]{0,0,0}$(u^*,v^*)$}%
}}}
\end{picture}

\]
We proceed to obtain bounds for all quantities at $(u^*,v^*)$,
independent of the choice of $(u^*,v^*)$.

Integrating $(\ref{rvqu})$ 
along $u=u^*$, in view of the inequality $\partial_vr=\lambda\ge0$
on $\mathcal{R}\cup\mathcal{A}$, we obtain
\[
r(u^*,v^*)=r(u^*,v_\epsilon)+
\int_{v_\epsilon}^{v^*}\lambda(u^*,v)dv\ge r_0,
\]
whereas integrating
$(\ref{ruqu})$ along $v=v^*$, we obtain
\[
r(u^*,v^*)\le R.
\]
Thus, the bounds $(\ref{yeniyildiz2})$ hold throughout
$D^+(X)\setminus\{p\}$.

We now proceed to show \emph{a priori} bounds for the mass:
\begin{equation}
\label{massfragmata}
-\frac{R^3C}3-M\le m\le \frac{R^3C}3+M
\end{equation}
throughout $D^+(X)\setminus\{p\}$.
Given $(u^*,v^*)$, 
we first show the right inequality of $(\ref{massfragmata})$. Since
$D^+(X)\setminus\{p\}\subset\mathcal{R}\cup\mathcal{A}$, 
we have--applying
inequalities $(\ref{duvamiko1})$, $(\ref{yeniyildiz})$, 
and $\nu<0$, to equation $(\ref{puqu})$--that 
\[
\partial_um\le r^2V(\phi)\nu\le -r^2C\nu,
\]
in this region.
Integrating the above inequality along $v=v^*$,
yields 
\[
m(u^*,v^*)\le m(u_\epsilon,v^*)+\int_{u_\epsilon}^{u^*}\partial_um\le M+\frac{R^3C}3
\]
as desired.
To show the left inequality of $(\ref{massfragmata})$ we 
proceed similarly, i.e.~from the inequality
\[
\partial_vm\ge r^2V(\phi)\lambda\ge -r^2C\lambda
\]
we obtain, integrating in $v$,
\[
m(u^*,v^*)\ge -\frac{R^3C}3-M.
\]
We have obtained, thus, that $(\ref{massfragmata})$ indeed holds throughout
$D^+(X)\setminus\{p\}$.

From this bound on $m$, it now follows by integrating $(\ref{pvqu})$,
and $(\ref{puqu})$ that we have
uniform \emph{a priori} integral estimates:
\begin{eqnarray*}
\left|\int_{u_\epsilon}^{u^*}{\frac12(1-\mu)\left(\frac\zeta\nu\right)^2\nu(u,v^*)du}
+\int_{u_\epsilon}^{u^*}V(\phi)r^2\nu(u,v^*) du\right|=\\
\left|\int_{u_\epsilon}^{u^*}{\frac12(1-\mu)\left(\frac\zeta\nu\right)^2\nu du}
+\int_{V(\phi)\ge 0}V(\phi)r^2\nu du
+\int_{V(\phi)<0 }V(\phi)r^2\nu du\right|\le\\
 \frac{2R^3C}3+2M
\end{eqnarray*}
and thus
\[
\int_{u_\epsilon}^{u^*}{\frac12(1-\mu)\left(\frac\zeta\nu\right)^2(-\nu)du}
+\int_{V(\phi)\ge 0}V(\phi)r^2(-\nu)du\le \frac{2R^3C}3+2M+\frac{CR^3}3.
\]
In view of the sign of $\nu$, and inequality $(\ref{yeniyildiz})$,
we have that both terms on the left are nonnegative. 
We obtain in particular
\begin{equation}
\label{kivntikn}
\int_{u_\epsilon}^{u^*}{\frac12(1-\mu)\left(\frac\zeta\nu\right)^2(-\nu)
(u,v^*)du}\le 
R^3C+2M.
\end{equation}
In an entirely similar fashion, we can obtain
\[
\int_{v_\epsilon}^{v^*}\frac12\kappa^{-1}\theta^2(u^*,v)dv\le
R^3C+2M.
\]
Integrating now $(\ref{vphi})$,
we obtain
\begin{eqnarray*}
|\phi(u^*,v^*)|	&\le&	|\phi(u^*,v_\epsilon)|
		+\left|\int_{v_\epsilon}^{v^*}{\frac\theta r(u^*,v)dv}\right|\\
	&\le&	P+\sqrt{\int_{v_\epsilon}^{v^*}
			{\theta^2\kappa^{-1}dv}}\sqrt{
			\int_{v_\epsilon}^{v^*}
			{\frac1{r^2}\kappa dv}}\\
	&\le&	P+2\sqrt{R^3C+M}\sqrt{r_0^{-2}K\epsilon}\\
	&=&	P_b.
\end{eqnarray*}

To estimate
the first derivatives of $r$ and $\phi$, let us rewrite
equations $(\ref{evolution1})$ and $(\ref{evolution3})$ as:
\begin{equation}
\label{nqu}
\partial_v\nu=\nu\left(2\kappa\left(\frac{m}{r^2}
-rV(\phi)\right)
\right),
\end{equation}
\begin{equation}
\label{sign1}
\partial_u\theta=-\frac{\zeta\lambda}r+\nu\kappa rV'(\phi),
\end{equation}
\begin{equation}
\label{sign2}
\partial_v\zeta=-\frac{\theta\nu}r+\nu\kappa rV'(\phi),
\end{equation}
and let us denote by $C'_b=\sup_{|x|\le P_b}{V'(x)}$
and $C_b=\sup_{|x|\le P_b}{V(x)}$. 
These constants are finite
in view of assumption $(\ref{duvamiko2})$.

Integrating
$(\ref{sign1})$,
we obtain
\begin{eqnarray*}
|\theta(u^*,v^*)|	&\le&	|\theta(u_\epsilon,v^*)|+
					\left|\int_{u_\epsilon}
						^{u^*}{\frac{\zeta\lambda}r(u,v^*)
					du}\right|+\left|
					\int_{u_\epsilon}^{u^*}
						{r\nu\kappa
					V'(\phi)du}\right|\\
			&\le&	\Theta+\left|\int_{u_\epsilon}^{u^*}{\frac\zeta\nu
						\kappa^{-1}
						\frac{(1-\mu)\nu}{r} du}\right|
						+R^2C'_bK\\
			&\le&	\Theta+\sqrt{\int_{u_\epsilon}^{u^*}
						{\left(\frac\zeta\nu\right)^2
						(-\nu)(1-\mu)du}
						\int_{u_\epsilon}^{u^*}
						{
						\kappa^2\frac{(-\nu)(1-\mu)}{r^2}
						du}}\\
			&&\hbox{}
						+R^2C'_bK\\
		&\le&	\Theta+2K\sqrt{2R^3C+2M}
			\sqrt{r_0^{-1}+\left(\frac{R^3C}3+
				M\right)r_0^{-2}}+R^2C'_bK\\
		&=&	\Theta_b.
\end{eqnarray*}
Integrating $(\ref{nqu})$ we obtain
\begin{eqnarray*}
-\nu(u^*,v^*)&\le&|\nu(u^*,v_\epsilon)|\exp{\left|\int_{v_\epsilon}^{v^*}
				2\kappa
				\left(\frac{m}{r^2}-rV(\phi)\right)dv\right|}\\
		&\le& N\exp{\left(2K\left(
				r_0^{-2}\left(\frac{R^3C}3+M\right)
				+RC_b\right)\epsilon\right)}\\
		&=&	N_b,
\end{eqnarray*}
while integrating $(\ref{nqu})$ in $u$, since $\partial_u\lambda=\partial_v\nu$,
we obtain
\begin{eqnarray*}
\lambda(u^*,v^*)&\le&\lambda(u_\epsilon,v^*)+
			2K\left(r_0^{-1}\left(\frac{R^3C}3+M\right)
				+R^2C_b\right)\\
		&\le&\Lambda+2K\left(r_0^{-1}\left(\frac{R^3C}3+M\right)
				+R^2C_b\right).
\end{eqnarray*}
Finally, integrating $(\ref{sign2})$, we obtain
that
\begin{eqnarray*}
|\zeta(u^*,v^*)|	&\le& Z+ \int_{v_\epsilon}^{v^*}
			\frac{|\theta\nu|}{r}(u^*,v)dv+\int_{v_\epsilon}^{v^*}
			\left|r\nu\kappa
			V'(\phi)(u^*,v)dv\right|\\
	&\le& Z+\Theta_bN_br_0^{-1}\epsilon+RC'_b\epsilon KN_b.
\end{eqnarray*}

We have estimated uniformly $|r^{-1}|$, $|\Omega^{-1}|$, 
$|\partial_ur|$, $|\partial_vr|$,
$|\phi|$, $|\partial_u\phi|$, and $|\partial_v\phi|$ in $D^+(X)\setminus\{p\}$.
By $(\ref{nqu})$, we have clearly also estimated $|\partial_u\partial_vr|$. Thus,
it remains only to estimate $\partial_u\partial_ur=\partial_u\nu$, 
$\partial_v\partial_vr=\partial_v\lambda$, $\partial_u\Omega$, $\partial_v\Omega$.

These estimates, it turns out, are now quite straight-forward.
Differentiating $(\ref{nqu})$ in $u$ we obtain,
\begin{eqnarray*}
\partial_v(\partial_u \nu)&=&
\partial_u \nu\left(2\kappa\left(\frac{m}{r^2}
-rV(\phi)\right)\right)\\
&&\hbox{}+\nu\left(2\partial_u\kappa\left(\frac{m}{r^2}
-rV(\phi)\right)\right)\\
&&\hbox{}+\nu\left(2\kappa\left(\frac{\partial_um}{r^2}-2\frac{m}{r^3}\nu
-\nu V(\phi)-V'(\phi)\zeta\right)\right).
\end{eqnarray*}
In view now of the bounds derived previously and $(\ref{para3})$, integrating
this equation in $v$
immediately yields
a uniform bound  
\[
|\partial_u \nu|\le \tilde{C}.
\]
We leave to the reader explicit calculation of the constant.
One argues similarly to obtain
\[
|\partial_v \lambda|\le \tilde{C}.
\]

On the other hand, integration of $(\ref{evolution2})$ in $u$ and
$v$ respectively, in view of the bounds derived previously and
the initial estimates $(\ref{para1})$ and $(\ref{para2})$,
gives uniform bounds 
\[
|\partial_v \Omega|\le \tilde{C},
\]
and 
\[
|\partial_u \Omega|\le \tilde{C}.
\]
Again, the details are left to the reader.

Thus, we
have shown that 
\[
N(D^+(X)\setminus\{p\})<\infty.
\] 
By Property~\ref{epekt}, we have
\[
p\in\mathcal{Q}^{>0}.
\] 
By continuity of $r$ and $m$, it follows
that 
\[
p\in\mathcal{R}\cup\mathcal{A}.
\]
$\Box$

\section{Remarks on the global structure of spacetime}
\label{globrem}

From the above, one easily proves
\begin{theorem}
\label{orizovtas9}
If $\mathcal{Q}^+\setminus J^-(\mathcal{I})\ne\emptyset$,
then there exists a null curve $\mathcal{H}^+\subset\mathcal{R}\cup\mathcal{A}$,
such that 
\begin{equation}
\label{orismosorizovta}
\mathcal{H}^+=
\overline{J^-(\mathcal{I})\cap\mathcal{Q}^+}
\setminus (I^-(\mathcal{I})\cup \overline{\mathcal{I}})
\end{equation}
\end{theorem}
\noindent\emph{Proof.}
Considet the set $\mathcal{H}^+$ defined by $(\ref{orismosorizovta})$.
Let $p\in\overline{\mathcal{Q}^+}$ denote the future limit endpoint of 
$\mathcal{H}^+\cap\mathcal{Q}^+$. In view of Proposition~\ref{inview},
we clearly have
\[
\mathcal{H}^+\cap\mathcal{Q}^+\subset \mathcal{R}\cup\mathcal{A}
\]
If $p\in\overline{\mathcal{I}}$,
then there is nothing to show.
If $p\not\in\overline{\mathcal{I}}$: 
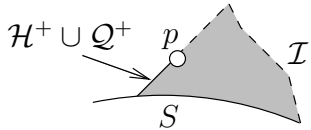
\[
\begin{picture}(0,0)%
\includegraphics{B.pstex}%
\end{picture}%
\setlength{\unitlength}{3947sp}%
\begingroup\makeatletter\ifx\SetFigFont\undefined%
\gdef\SetFigFont#1#2#3#4#5{%
  \reset@font\fontsize{#1}{#2pt}%
  \fontfamily{#3}\fontseries{#4}\fontshape{#5}%
  \selectfont}%
\fi\endgroup%
\begin{picture}(1852,833)(2961,-3957)
\put(4726,-3511){\makebox(0,0)[lb]{\smash{\SetFigFont{12}{14.4}{\rmdefault}{\mddefault}{\updefault}{\color[rgb]{0,0,0}$\mathcal{I}$}%
}}}
\put(3936,-3379){\makebox(0,0)[lb]{\smash{\SetFigFont{12}{14.4}{\rmdefault}{\mddefault}{\updefault}{\color[rgb]{0,0,0}$p$}%
}}}
\put(2961,-3399){\makebox(0,0)[lb]{\smash{\SetFigFont{12}{14.4}{\rmdefault}{\mddefault}{\updefault}{\color[rgb]{0,0,0}$\mathcal{H}^+\cup\mathcal{Q}^+$}%
}}}
\put(3908,-3908){\makebox(0,0)[lb]{\smash{\SetFigFont{12}{14.4}{\rmdefault}{\mddefault}{\updefault}{\color[rgb]{0,0,0}$S$}%
}}}
\end{picture}

\]
then it follows
easily that $p$ is a first singularity. Clearly, $p\not\in\overline{\Gamma}$,
i.e.~$(\ref{suv9nkn1})$ does not hold.
Since $J^-(p)\cap\mathcal{Q}^+\subset \overline {J^-(\mathcal{I})}$,
it follows from Proposition~\ref{inview}
 that $J^-(p)\cap\mathcal{Q}^+\subset \mathcal{R}\cup\mathcal{A}$.
 Thus, $(\ref{suv9nkn2})$ does not hold either. We contradict
 the statement of Theorem~\ref{kurio}. $\Box$

Note that in view of Corollary~\ref{allo}, a sufficient condition for
the assumption of Theorem~\ref{orizovtas9} is $\mathcal{A}\cup\mathcal{T}\ne\emptyset$.
Moreover, in view of Theorem~\ref{kurio},
a sufficient condition for $\mathcal{T}\ne\emptyset$
is that there exists a first singularity $p$ 
such that $p\not\in\overline\Gamma$,
in particular, this is the case if ``a component of the singularity is spacelike''.
Thus, it is clear that the naked singularities, as described 
in~\cite{hhm:gccv}, can in fact never arise.

In general, we note that 
this argument says nothing about the behaviour
of $r$ on the event horizon. In fact, \emph{a priori} we could have
$r\to\infty$, i.e.~$\mathcal{H}$ could have a limit point on $\mathcal{I}$
itself. On the other hand,
in the case of a Higgs field with \emph{non-negative} 
potential\footnote{in this case, the positive energy condition holds}
 collapsing from spherically symmetric asymptotically flat initial data, 
we can say more: Indeed, the results of this paper clearly
imply that such fields satisfy the assumptions of~\cite{md:sssts}.
It then follows from the results of~\cite{md:sssts}
that if $\mathcal{Q}^+\setminus J^-(\mathcal{I})\ne\emptyset$, 
then null infinity is complete (see below), and moreover
a Penrose inequality holds bounding the area radius of
the event horizon by twice the final Bondi mass, which is in turn finite.

\section{Weak cosmic censorship?}
\label{wcc??}
Weak cosmic censorship is the conjecture that for generic initial data,
$\mathcal{I}$ is complete\footnote{Completeness can be defined 
with respect to an appropriate induced connection on $\mathcal{I}$ related to 
the conformal compactifications given by the Penrose diagrams 
we are employing. For a definition
of the notion of completeness
applicable in the general (not necessarily spherically symmetric)
asymptotically flat case, without \emph{a priori} regularity 
assumptions at infinity, see~\cite{chr:givp}.}. 
We have noted above that in 
the case $V\ge 0$ and $\mathcal{Q}^+\setminus J^-(\mathcal{I}^+)\ne\emptyset$,
the completeness of $\mathcal{I}^+$ in the asymptotically
flat setting follows from~\cite{md:sssts}. We remark, here, 
that in the case where it is only assumed that $V\ge -C$, 
then \emph{a priori}, infinity may be either complete or incomplete,
despite the fact that
$\mathcal{H}$ is regular. Understanding the global 
properties of these spacetimes, thus, requires
further examination.

\section{Acknowledgement}
The author thanks Gary Horowitz, Mukund Rangamani, and David Garfinkle
for helpful discussions on a previous version of this paper. 
Since the original appearance of this paper on the gr-qc arxiv, the scenario
of~\cite{hhm:gccv} has been updated, so as--in particular--not to be
inconsistent with the results
proven here. See~\cite{hhm:update}  and references therein.

\end{document}